\documentclass[pre,twocolumn,showpacs,amsmath,amssymb,superscriptaddress]{revtex4}
\usepackage{graphicx,amstext}
\usepackage{dcolumn}
\usepackage{bm}
\newcommand{\vs}{{\it vs.\@}}
\newcommand{\ie}{{\it i.e.\@}}
\newcommand{\al}{{\it et al.\@}}
\newcommand{\bq}{\begin{equation}}
\newcommand{\eq}{\end{equation}}
\begin{document}

\title{Intermediate DNA at low added salt: DNA bubbles slow the diffusion of short DNA fragments }

\author{T.~Vuleti\'{c}}
\homepage{http://tvuletic.ifs.hr/}
\email{tvuletic@ifs.hr}
\affiliation{Institut za fiziku, 10000 Zagreb, Croatia} 
\author{S.~Dolanski Babi\'{c}}
\affiliation{Institut za fiziku, 10000 Zagreb, Croatia}
\altaffiliation {Permanent address: Department of physics and biophysics,
Medical School, University of Zagreb, 10000 Zagreb, Croatia.}
\author{T.~Ban}
\affiliation{Institut za fiziku, 10000 Zagreb, Croatia}
\author{J.R\"{a}dler}
\affiliation{Ludwig-Maximilians-Universit\"{a}t, Sektion Physik,
Geschwister-Scholl-Platz 1, D-80539 Munich, Germany}
\author{F.~Livolant}
\affiliation{Laboratoire de Physique des Solides, Universit\'{e} Paris Sud -
F-91405 Orsay, France}
\author{S.~Tomi\'{c}}
\affiliation{Institut za fiziku, 10000 Zagreb, Croatia}
\date{\today}

\begin{abstract}

We report a study of DNA (150 bp fragments) conformations in  very low added 
salt $<0.05$mM, across wide DNA concentration range $0.0015\leq c \leq 8$~mM 
(bp).   We found an intermediate DNA conformation in the region $0.05 < c < 1$~mM, by means 
of fluorescence correlation spectroscopy (FCS) and UV-absorbance measurements. 
FCS detected that in this region DNA has the diffusion coefficient, $D_p$ 
reduced below the values for both ssDNA coils and native dsDNA helices of 
similar polymerization degree $N$. Thus, this DNA population can not be a 
simple mix of dsDNA and of ssDNA which results from DNA melting. Here, melting 
occurs due to a reduction in screening concomitant with DNA 
concentration being reduced, in already very low salt conditions. The 
intermediate DNA is rationalized through the well known concept of fluctuational 
openings (DNA bubbles) which we postulate to form in AT-rich portions of 
the sequence, without the strands coming apart. Within the bubbles, DNA is 
locally stretched, while the whole molecule remains rod-like due to very low salt environment. Therefore, such intermediate DNA is elongated, in comparison to dsDNA, which accounts for its reduced $D_p$.  

\end{abstract}


\maketitle

\section{Introduction}
\label{sec1}
For transcription to proceed in a
chromosome, DNA has to be separated from the histone proteins and then it has to
be unwound and the strands separated to allow the access of transcription
factors (enzymes). This rather complicated process has, from a physicist  point
of view, a lot in common with a simple, thermally induced, {\em in vitro}
process of DNA melting/denaturation, where dsDNA helix completely separates into
two ssDNA strands. Stability of the double stranded DNA molecule reflects the
nature of the DNA as a polyelectrolyte \cite{Polymerbooks,Bloomfield00}. When
dissolved in polar solvents polyelectrolytes dissociate into a polyion and small
counterions. Electrostatic interactions and entropy of all those charges, as
well as those of added salt control their phenomenology.  Thus, the stability of
dsDNA depends on added salt, DNA concentrations and temperature
\cite{Dove62,Record75,Chen93,Bishop}. The DNA melting temperature will decrease with
a decrease in added salt, and also with a decrease in concentration of DNA
itself, even more so if there is no added salt. The latter is due to a change in
concentration of counterions.  {\em E.g.,} at high added salt, up to
physiological levels of the order of 0.1-1 M the DNA would melt at high
temperatures (80-97$^{o}$C). Contrary, below 1 mM (total counterions and added
salt) dsDNA may be unstable already at ambient temperature
\cite{Record75,Auer69}. 

These extrinsic parameters are complemented with
intrinsic ones: the DNA sequence, conformation and structure. Primarily, the
sequence of A-T or G-C pairs influences the melting. For example, G-C base pairs
are connected with three hydrogen bonds, while the A-T pairs with only two, and
thus the sequences rich in A-T pairs are less thermally stable.  The stability
of a Watson-Crick pair is also determined by its first neighbours, due to
stacking interactions \cite{Breslauer86}. This dependence on the local sequence
relates to a rather well accepted concept \cite{Yakusevich04} that the
denaturation proceeds from local, cooperative openings of several bp long, A-T
rich sequences, called DNA bubbles, which occur well below melting temperature
\cite{Frank87,Gueron87,Frank06}. Close to the  melting temperature these
fluctuations {\em zipper} along the strands and the strands separate. The
bubbles indeed correlate with the actual transcription initiation sites
\cite{Rapti06}. Physically, there is about a 10 $k_BT$ barrier for opening a
bubble and the cost for breaking consecutive base-pairs is only about 0.1
$k_BT$, which results in the so-called {\em zippering} denaturation process
\cite{Manghi07}. For the very short fragments, $\approx 10$ bp, the transition
may well be described with a two-state model. For a longer DNA molecule,
partially open states may be conceived,  where a part of the sequence is open
and part still closed. Such molecules, if stable, may be recognized to be the
intermediate DNA, nor ssDNA, neither dsDNA \cite{virus,Bishop05,Peyrard09JPCM}.

Models for DNA denaturation start from these general concepts. Models may be
simple Ising-like, starting from the original Poland-Scheraga model
\cite{Scheraga66}, where only the energy difference is counted for open and
closed basepairs. In more detailed models base interaction depends on the
distance between the bases (Dauxois-Peyrard-Bishop type)\cite{Bishop}. Some
models account for the entropy contribution from the more flexible ssDNA loops
\cite{Manghi07,Manghi09}, and find a continuous (II order) or discontinuous (I
order) transition \cite{Kafri00,Carlon02}. The latter result was crucially based
on taking into account the self-avoiding interactions between the various parts
of the chain.

For the experimentalist, it is not straightforward to demonstrate the existence
of the DNA bubbles. Certainly, experiments evaluating the basepair opening
probability rates and bubble lifetimes, in any specific set of conditions for
their existence, would contribute to the modeling effort and our general
knowledge of DNA stability and dynamics. The information on base-pair opening
was initially provided by NMR study \cite{Gueron87,Gueron88} of the imino-proton
exchange, \ie~study of protons, formerly participating in hydrogen bonds, that
bases exchange with D$_2$O water when a base pair opens. This NMR study actually
points that the bubble opening is a very local phenomenon, occuring on the level
of a single base-pair. Opening probabilities (dissociation constants) at 25$^o$C
are actually quite low: $10^{-5}-10^{-6}$ for an AT pair  and $10^{-6}-10^{-7}$
for a GC pair. The lifetime of a basepair is of the order of 10 ms and the rate
for closing the bubble is $10^{8}$s$^{-1}$. Only recently these NMR studies have
been complemented by other studies of bubble dynamics. An FCS study by Libchaber
\al~\cite{Libchaber03} has found that fluctuations  occur with closing rates 3-4
orders of magnitude higher than estimated by NMR, {\em i.e.}
$10^{4}-10^{5}$s$^{-1}$. The size of the bubbles was 2 to 10 bp on a fragment
only 18bp long. These authors found no contradiction with NMR rate estimates,
concluding that NMR registers much smaller conformational change with respectively higher fluctuation rates. That is, they considered
that the openings detected by FCS are more biologically relevant as they  relate
to larger, nm-scale separations.

Peyrard \al~\cite{Peyrard09JPCM,Peyrard09EPL} were able (by two-photon
excitation of guanines) to  detect when GC pairs were not paired in short
$\approx50$bp DNA while temperature was raised and DNA melted. Thus they
obtained localized information on bubble opening and provided sequence dependent
DNA melting curves. Importantly, they noted that short AT rich sequences (TATA
boxes) would fluctuate and disturb both neighbouring and distant GC pairs well
below the melting temperature. The fact that the opening fluctuations can have a
non-local (non-single basepair) effect is related to nonlinear dynamics of DNA
\cite{Yakusevich04}, but it was emphasized that this effect becomes negligible
as it averages out for kbp or longer DNA sequences.

From the above it may be recognized that intermediate DNA states should exist
below the melting transition, however only one group of authors has insofar
managed to quantify the fraction of intermediate DNA in the population of DNA
molecules. Montrichok \al~\cite{Montrichok03} have designed DNA constructs, ones
with the TATA box at one end, and the others with TATA box ("soft" region) in
the middle and GC clamps at both ends. When melting was stopped before complete
denaturation, and the solution cooled quickly (quenched), strands which were
completely separated were not able to find their complements and formed hairpins
instead. Their fraction was quantified by simple gel electrophoresis. Fraction
of the open molecules was compared (at equal temperatures) to the fraction of
open basepairs in the solution, quantified by UV-absorption measurements. If the
latter fraction was higher then there must have been some partly denatured,
intermediate molecules before the quench. Intermediate DNA states identified
from these experiments were succesfully modeled by Monte Carlo simulations using
the Dauxois-Peyrard-Bishop model \cite{Bishop05}.

One issue that arises is whether the local fluctuations influence mechanical 
properties of the whole DNA molecule. Early experiment, where self-diffusion 
coefficient of 17 kbp DNA molecules was probed in temperature by dynamic light 
scattering found no effect of fluctuations on DNA rigidity (persistence length) 
\cite{Wilcoxon83}. Only recently it has been noted that there is the effect of 
fluctuations, but only on very short (10 -90 bp) DNA fragments, which could 
behave as having a persistence length as low as 10-20 nm \cite{Yuan08,Lee10}, 
much shorter than the agreed value $L_p=50$ nm \cite{Baumann97}. Also, Tomi\'{c} 
\al~ have noted that persistence length of DNA might be reduced in some 
conditions~\cite{TomicEPL08}. They worked with mononucleosomal $\approx150bp$ 
DNA fragments, which have the contour length $L_c=50$ nm similar to the 
persistence length, so they are practically rodlike. Their dielectric spectroscopy technique, which provides 
characteristic length scales of a polyelectrolyte system, complementary to {\em 
e.g.} data by small angle X-ray scattering \cite{ESOPS}, indeed found a 50 nm 
length scale for fragments in pure water. However for fragments in 1 mM NaCl the 
characteristic length scale was 20-30 nm, lower both than $L_c$ or $L_p$. This 
would indicate that added salt, which does reduce the electrostatic contribution 
to the persistence length \cite{OSF}, has revealed $L_p$ lower than the agreed 
value. We may presume that the reduced $L_p$ is the consequence of partial DNA 
denaturation in thesee conditions of low salt.

In this paper, we propose that intermediate DNA is the major constituent in 
mononucleosomal DNA (denoted as DNA146) conformations population in the 
concentration range $0.05 < c < 1$~mM (bp) at very low salt  conditions, 
$c_{\mathrm{salt}}<0.05$mM. We infer the presence of intermediate DNA from the 
DNA146 polyion self-diffusion coefficient $D_p$, obtained by the fluorescence 
correlation spectroscopy  measurements (FCS, method described in Sec.II).  In 
Sec.III we present the $D_p$ value for $0.05 < c < 1$~mM range which is below 
the values for both ssDNA and native dsDNA. This means that DNA conformation 
population there can  not be a simple mix of native dsDNA helix and separated 
ssDNA strands. In Sec.IV we argue that an intermediate DNA conformation must be 
present, hydrodynamically larger than both ssDNA or dsDNA of similar 
polymerization degree. We suggest that under these conditions the basepair 
opening and closing rates have changed in such a manner to induce a variation of 
the diffusion properties of the whole molecule. In this sense, the intermediate 
DNA may be considered to be a stable form.  Eventually, we analyze $D_p$ data 
and extract the fraction of completely open DNA molecules, ssDNA, against partly 
open, intermediate DNAs and  native dsDNA. We also  compare these fractions with 
measured fraction of open DNA base-pairs, following the procedure of Montrichok \al~\cite{Montrichok03}.

\section{Materials \& Methods}
\subsection{Samples}
We studied solutions of nucleosomal dsDNA fragments about 150 bp and 50 nm long.
This DNA is denoted as DNA146, since there are 146 bp of DNA wrapped around a
histone octamer in a nucleosome core particle. DNA146 pellets from the same
stock as in \cite{previous} was used. The monodispersity of fragments was
checked by gel electrophoresis. We also checked whether the pellets contain any
added salt. DNA concentrations will be expressed as molar concentrations of
basepairs in the remainder of this paper.

Several solution sets with different DNA146 and added salt concentrations were
prepared, which we describe below:
\begin{itemize}
\item[set I] {\em Solutions with varying DNA concentration, in very low added
salt:}
A set of 40 solutions covering the range of DNA concentrations $c=0.001-8$mM was 
prepared by dilution with pure water of aliquots from the  mother solution to 
reach the desired concentration. The mother solution was obtained by dissolving 
10 mg of dry DNA146 in 0.55 mL pure water - the final concentration was $c=27$ 
mM. The UV-absorbance at 260 nm was measured, at 25$^o$C for all the solutions 
in the set. Only then, for FCS measurements, the DNA146 solutions were 
fluorescently labeled by addition of synthetic 110bp DNA from a 0.5 mM stock 
kept in 10 mM TE buffer. This 110bp dsDNA will be further denoted DNA110*. The 
asterisk * denotes the fluorescent labeling by Cy5 fluorophore at one end of DNA 
molecule \cite{previous}. DNA110* was added to a concentration 1.5-2 $\mu$M to 
achieve 20 nM Cy5 label concentration. The buffer of DNA110* stock gets diluted 
after addition into pure water DNA146 samples, thus introducing a minimal added 
ionic strength, not above 0.05mM. In this manner we regarded these samples as 
very low salt solutions, $c_{\mathrm{salt}}<0.05$mM, and labeled them 
appropriately. FCS results for this set are presented also in \cite{previous}.

\item[set II] {\em Solutions with varying DNA concentration, in 10 mM TE buffer}:
From $c=27$ mM pure water mother solution (see above) a 0.1 mL sample was taken 
and diluted with 0.25 mL of 35 mM TE buffer. We used 10:1 TrisHCl:EDTA buffer 
set to pH7.5. Thus we obtained DNA146 starting solution with $c=7.7$ mM in 10 mM 
TE buffer .  For FCS measurements DNA110* was added to this solution, to a 
concentration 1.5-2 $\mu$M. After FCS measurement the sample was further diluted 
with appropriate volume of 10 mM TE buffer to produce a next lower DNA146 
concentration sample. The procedure was repeated and 15 different DNA 
concentrations were measured in the range $c=0.03-8$mM.  On each dilution 
appropriate volumes of DNA110* stock were being added to maintain the 20 nm Cy5 
level.

\item[set III] {\em Solutions with fixed DNA concentration, in varying added
salt:}
DNA146 mother solution, $c=7.5$ mM was prepared by dissolving a pellet in 10 mM
NaCl. By dilution with appropriate volumes of pure water and 1mM NaCl we
prepared a solution set where DNA concentration was always $c=0.75$ mM, while 
NaCl concentrations were 0.2, 0.5, 1, and 2 mM.
\end{itemize}

\subsection {UV absorption measurements }
DNA146 sample set I  (varying DNA, very low salt $<0.05$ mM) has 
been tested without dilution for UV-absorbance at 260 nm with Nanodrop 
(ThermoScientific). This instrument shows a linear response in absorbance for a 
very broad DNA concentration range, $c=0.003-15$~mM, with the sample volume of 
only 1-2 $\mu$L.  The Nanodrop instrument was critical to get the actual UV 
absorbance of DNA samples without dilution.

For DNA melting experiments on set III (fixed DNA, varying salt) in 0-90 $^{o}$C 
range we have used Agilent - HP 8452 Spectrophotometer with a Peltier based 
temperature control. From the temperature dependent UV-absorbance at 260 nm 
(hyperchromicity) curves we extracted the melting transition temperature $T_m$ 
and transition width dependence on total ion concentration.

\subsection{Fluorescence correlation spectroscopy}
We have used Zeiss ConfoCor II FCS instrument. Focal volume was defined by a 
Zeiss Plan-NeoFluar 100x/NA1.3 water immersion objective, epi-illumination was 
by He-Ne 632.8 nm 5mW laser, for excitation of Cy5 fluorophore. Measurements 
were performed at $25^{o}$C, the ambient temperature of the temperature 
stabilized clean-room.

Fluorescence correlation spectroscopy is used to measure the diffusion 
coefficient of the fluorescently labeled molecules. Number fluctuations of the 
molecules entering and leaving the focal volume of the instrument are registered 
as fluorescence variation. An autocorrelation function is calculated for the 
fluorescence intensity trace.   This function  decays exponentially with 
autocorrelation time $\tau_c$:

\begin{equation}
G(\tau_c) = \frac{1}{N_f} \cdot \frac{1} {1 +\frac{\tau_c}{\tau}}
\frac{1}{\sqrt{(1 + (\frac{w_0}{z_0})^2\frac{\tau_c}{\tau})}} (1 +
\frac{T}{1-T}exp(-\frac{\tau_c}{\tau_T}))
\label{FCSautocorr}
\end{equation}
$N_f$ is the average number of fluorescent molecules in the focal volume, 
$z_0/w_0$ is the focal volume structure parameter and $T$, average fraction of 
fluorophores in the triplet state (thus, non-fluorescing). The lifetime $\tau_ 
T$ of the triplet state is taken into account when fitting. The characteristic 
decay time $\tau$ is the diffusion time that the fluorescent molecule takes to 
traverse the focal volume. The details of the procedure we used to extract 
$\tau$ may be found in \cite{previous}.

\begin{figure}
\resizebox{0.46\textwidth}{!}{\includegraphics*{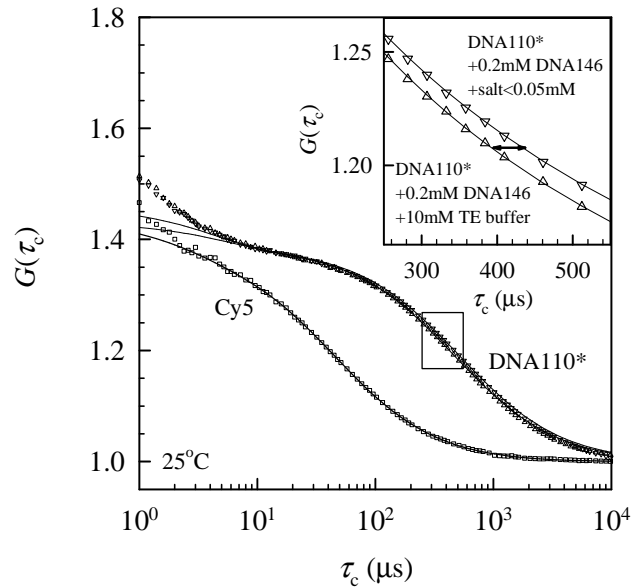}}
\caption{
Experimental autocorrelation functions $G(\tau_c)$ (denoted DNA110*) recorded for a sample from 
set I (2 $\mu$M DNA110* in 0.2 mM DNA146 in very low salt $<0.05$mM) or a sample 
from set II (2 $\mu$M DNA110* in 0.2 mM DNA146 in 10 mM TE buffer). For comparison, the autocorrelation function (denoted Cy5) 
is shown for 20 nM Cy5 fluorophore in pure water, without any DNA. The inset zooms the 
correlation time range $250-550 \mu$s, denoted by a rectangle in the main panel. 
The experimental values of $G(\tau_c)$ are shown with symbols and the respective 
fits with lines.  
}
\label{FCSraw}
\end{figure}

In Fig.\ref{FCSraw}, besides the autocorrelation function experimental curve and 
fit for Cy5, we present the $G(\tau_c)$ curves recorded for  2 $\mu$M of DNA110* 
found in either 0.2 mM DNA146  in very low salt $c_{\mathrm{salt}}<0.05$mM,  
(set I) or 0.2 mM DNA146 in 10 mM TE buffer (set II). The experimental curves 
have been normalized, and thus they appear to overlap  due to a rather small 
difference in the diffusion times. Nevertheless, the diffusion 
times $\tau$ that we extracted are different, with values of 390 and 430 $\mu$s, 
respectively.  To demonstrate that the fits reliably distinguish between the two 
experimental data sets for DNA110*, the area of the main panel, denoted with the 
rectangle, is shown enlarged in the inset \cite{Mangenot03}.  The arrow in the 
inset denotes the 40 $\mu$s difference between the respective diffusion times 
obtained by fits.

The diffusion time is inversely proportional to the self-diffusion coefficient 
$D_p$ of the molecule. The relationship between $\tau$ and $D_p$ may be obtained 
from the measurement of $\tau_{Cy5}$ for Cy5 fluorophore, whose diffusion 
coefficient is known, $D_{Cy5}=3.16 \cdot 10^{-10}$ m$^2$/s \cite{ZeissManual}:

\begin{equation}
D_p=D_{Cy5}\frac{\tau_{Cy5}}{\tau}
\label{DDCy5}	
\end{equation}
The diffusion times $D_p$ that we can obtain from $\tau$ are for the 
fluorescently labeled DNA110* molecules at a relatively low concentration, 
diffusing in solutions of varying, but mostly higher DNA146 concentration. That 
is, we obtain $D^{exp}_{110*}$, diffusion coefficient for DNA110* which, 
however, depends on DNA146 concentration $c$. However, we assume that it is 
possible to rescale $D^{exp}_{110*}(c)$ to get the values for DNA146 
$D^{exp}_{146}(c)$.  We remind that dsDNA persistence length $L_p=50$ nm 
\cite{Baumann97} is comparable to contour lengths of both DNA110* and DNA 146, 
38 and 50 nm, respectively. Therefore, we expect that they assume an extended, 
rodlike configuration. According to Tirado \al  \cite{Tirado84} the 
translational diffusion coefficient calculated for a rodlike macromolecule is 
given by

\begin{equation}
D^{th}=\frac{kT}{3\pi\eta}\frac{\ln(L_c/d)+0.312}{L_c}
\label{TiradoFormula}	
\end{equation}
Here $L_c=Nb$ is contour length, $d$ is polyion diameter, $\eta=8.9 \cdot 
10^{-4}$Pas is viscosity of water  ($T= 298$~K). With $b=0.34$ nm and $d=2.6$ 
nm, the diffusion coefficient for 110bp DNA is $D^{th}_{110*}=3.98 \cdot 
10^{-11}$ m$^2$/s and for 146bp DNA is $D^{th}_{110*}=3.27 \cdot 10^{-11}$ 
m$^2$/s. Stellwagen \al \cite{Stellwagen03} have reviewed the literature and 
shown that the expression by Tirado \al~is well applicable to experimental data 
obtained for DNA molecules in size from 10 to 1000 basepairs.

Thus,
\begin{equation}
D^{exp}_{146}(c)=\frac{D^{th}_{146}}{ D^{th}_{110*}} D^{exp}_{110*}(c)
\label{extrapol}	
\end{equation}
In this manner, fluorescence correlation spectroscopy was used to obtain the 
self-diffusion coefficient of DNA146, $D^{exp}_{146}(c)$ for varying DNA 
(DNA110* and DNA146 combined) concentrations $c=0.0015-8$ mM. Most importantly, 
we were able to distinguish and compare the values obtained for samples in very 
low salt (set I) and samples in 10 mM TE buffer (set II).

\section{Results}
\label{sec3}
\subsection{DNA melting in very low salt conditions}

Classically, UV absorbance measurements at 260 nm are used to study the DNA 
melting, \ie helix-coil transition \cite{Bloomfield00,Record75}. Since the 
extinction coefficient for the dsDNA, 1 O.D.=50 $\mu$g/mL is about 40\% higher 
than for ssDNA, 1 O.D.=35 $\mu$g/mL, concomitantly with temperature induced DNA 
denaturation, occurs a 40\% rise in absorbance (hyperchromicity effect).

\begin{figure}
\resizebox{0.46\textwidth}{!}{\includegraphics*{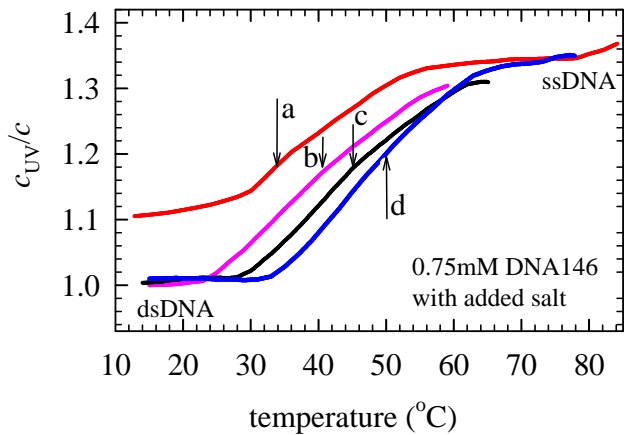}}
\caption{
Melting curves for DNA146, $c=0.75$~mM (bp)  with varying added 
salt 0.2, 0.5, 1, and 2 mM NaCl  (set III), denoted as a, b, c, d, respectively. 
Melting appears as an increase in  
$c_{UV}/c$ ratio. The arrows denote 
melting temperatures $T_m$ defined at half height of the maximum $c_{UV}/c$ rise.}
\label{UVtemp}
\end{figure}

In Fig.\ref{UVtemp}, we present thermal denaturation curves for DNA146 samples at a 
fixed DNA concentration  $c=0.75$ mM, in varying added salt, 0.2-2 mM NaCl. The measured UV 
absorbance rise is presented as the rise of the ratio between UV-spectrophotometrically 
determined concentration $c_{UV}$ and the gravimetrically determined (during 
sample preparation) concentration. The absorbance was converted into 
$c_{UV}$ concentrations using the extinction coefficient for the dsDNA. $c_{UV}/c$ ratio close to 1.4 thus corresponds to fully denatured DNA and $c_{UV}/c=1.0$ corresponds to the native dsDNA. Melting 
temperatures $T_m$ are therefore defined at half height of the maximum $c_{UV}/c$ rise, where half of the basepairs should be open. For DNA146 in 
0.2 mM NaCl solution  we note that already at ambient temperature UV-absorbance, ($c_{UV}/c>1$) is enhanced. Therefore we presume 
that this sample was already partly denatured at ambient conditions, due to the 
low salt content. 

\begin{figure}
\resizebox{0.46\textwidth}{!}{\includegraphics*{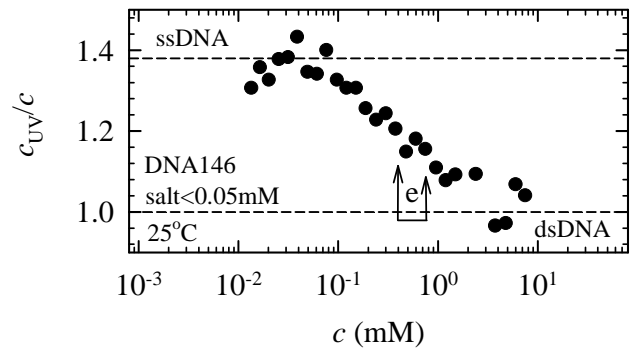}}
\caption{
The $c_{UV}/c$ ratio for DNA146 in very low salt 
$c_{\mathrm{salt}}<0.05$~mM (set I).  The dashed 
lines at $c_{UV}/c\approx1.4$ and 1 denote fully denatured ssDNA and native dsDNA, respectively. The arrows and letter $\mathrm e$  denote the concentration range where 
the ratio $c_{\mathrm{UV}}/c$ is just below 1.2, halfway between ssDNA and dsDNA 
levels.
}
\label{UVgrav}
\end{figure}

In Fig.~\ref{UVgrav} we present the ratio $c_{UV}/c$, measured at 25$^o$C for DNA146 dissolved in very low added 
salt $c_{\mathrm{salt}}<0.05$~mM (set I).  At higher concentrations, above $c=1$~mM $c_{UV}$ and $c$ seem to comply 
indicating that the DNA remains native. At the lowest concentrations $c_{UV}/c$ increases to about 1.4, which easily relates to 40\% hyperchromicity of denatured 
DNA. Therefore,  in Fig.~\ref{UVgrav} dashed 
lines at $c_{UV}/c\approx1.4$ and 1 are respectively labeled ssDNA and dsDNA. The 
scatter in the data is mostly due to errors in DNA sample preparation, and to 
some extent due to the very minute sample droplets applied (a cuvette free 
spectrophotometer is used), which may evaporate and change concentration rather 
quickly.

Further, we note DNA146 samples from the intermediate concentration range 
$c=0.4-0.75$ mM  that show $c_{UV}/c\approx1.2$, corresponding to about 20\% 
hyperchromicity ({\em cf.} Fig.\ref{UVtemp}). These samples are  thus  halfway through the melting 
transition, at ambient temperature. Thus, for DNA146 in the $c=0.4-0.75$ mM range we estimate the temperature where we performed measurements to be the melting temperature, $T_m=25^{o}$C. 

\begin{figure}
\resizebox{0.46\textwidth}{!}{\includegraphics*{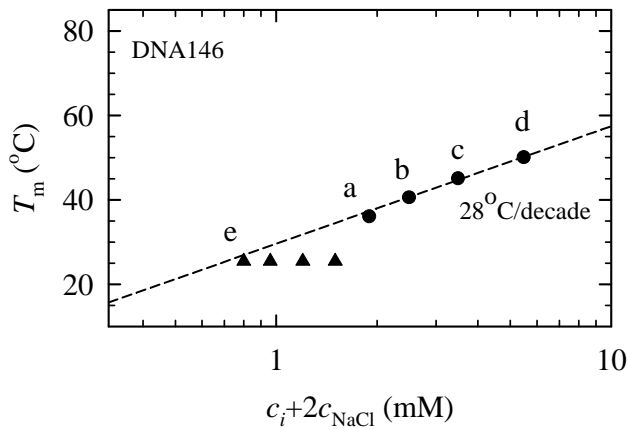}}
\caption{
DNA146 melting temperatures $T_m$, as a function of total ionic concentration 
$c_{i}+2c_{\mathrm{NaCl}}$ of the solution. Counterion concentration  $c_{i}$ is 
defined as twice the DNA basepair concentration $c$. Circles denote $T_m$ data 
obtained for $c=0.75$~mM DNA146 in varying salt (points $\mathrm {a,b,c,d}$, set III) solutions. For DNA146 
$c=0.4-0.75$~mM samples in very low salt (points $\mathrm e$, set I) we have estimated $T_m=25^o$C 
(triangles).
}
\label{phasdiag}
\end{figure}

In Fig.\ref{phasdiag} we show DNA146 melting temperature $T_m$ \vs~total ionic 
concentration of the solutions. We show the 
values obtained from melting curves  (sample set III, Fig.\ref{UVtemp}, points $\mathrm {a,b,c,d}$) alongside 
the just above defined $T_m=25^{o}$C for DNA146 $c=0.4-0.75$ mM concentration 
range (sample set I, Fig.\ref{UVgrav}, points $\mathrm {e}$). Total ionic concentration $c_{tot}$ is a sum of all the 
ions present in the solution: counterions $c_i=2c$ and added salt anions and 
cations $2c_{\mathrm{NaCl}}=c_{\mathrm{Na^+}}+c_{\mathrm{Cl^-}}$, thus $c_{tot} 
=c_{i}+2c_{\mathrm{NaCl}}$.

As predicted by Manning \cite{Manning72_a} we obtain the linear slope $d T_m/d 
\log c_{tot}=28\pm 3^{o}$C per decade, which compares very well with the slope 
27$^{o}$C per decade found by Record \cite{Record75} for T4 (170 kbp) and T7 (40 
kbp) bacteriophage DNA solutions without added salt. The slope reflects the 
counterion condensation and should be independent of the chain size and thus 
invariant from the type of DNA used in the experiment, as, indeed, we observed. 
While the slope is similar, we note that the transition temperatures for DNA146 
are about 10$^{o}$C lower than for long T4 and T7 DNAs, for equal $c_{tot}$. We 
also note that the melting transitions for DNA146 are much broader, 
$\approx$20-30$^{o}$C, than for T4 and T7 DNAs, where they are only 4-9$^{o}$C. 
That the melting temperature decreases and the transition width increases for 
shorter DNA chain size is to be expected \cite{Benight,Buyuk}.

The above determined ability of DNA146 to keep the native conformation at 
ambient temperature, in the very low salt ($<0.05$ mM) environment, but only 
above a certain DNA concentration $c=1$~mM, reflects the findings by Record 
\cite{Record75} and by Tomic \al~\cite{TomicPRE07}. In the following we identify  
the DNA conformations present in the very low salt solutions, with the region 
below $c=1$~mM being the most intriguing.

\subsection{DNA146 diffusion coefficient}

\begin{figure}
\resizebox{0.46\textwidth}{!}{\includegraphics*{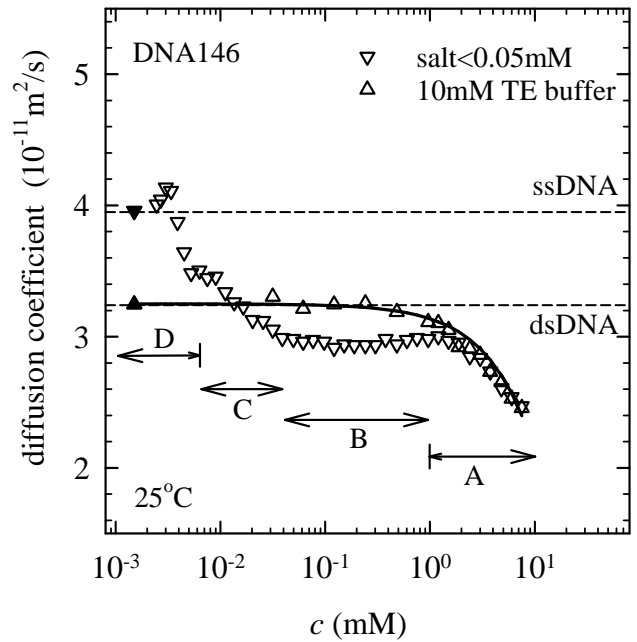}}
\caption{
Diffusion coefficient $D^{exp}_{146}$ for DNA146 in very low salt $<0.05$ mM 
(inverted triangles, sample set I) and in 10 mM TE buffer (triangles, sample set 
II). $D^{exp}_{146}$ varies with the DNA concentration $c$ (basepair) across 
regions denoted A to D. Full line denotes the fit to Eq.\ref{Dhont}, which 
describes the concentration dependence of diffusion coefficient for rod-like 
particles. Lower dashed line shows the theoretical diffusion coefficient value 
$D^{th}_{146}=3.27 \cdot 10^{-11}$ m$^2$/s  for native DNA146 helices in dilute 
solution. It also points to the value $D^{ds}_{146}$ extrapolated from the 
diffusion time measured for DNA110* in 10mM TE buffer, without DNA146 (see 
text). Upper dashed line points to the diffusion coefficient value (inverted 
black triangle)$D^{ss}_{146}=3.95 \cdot 10^{-11}$ m$^2$/s expected (see text) 
for 146 nt ssDNA strands in pure water.
}
\label{diffcoef}
\end{figure}

In Fig.~\ref{diffcoef} we present $D^{exp}_{146}$, the diffusion coefficient of 
DNA146 polyion as a function of DNA concentration $c$ in the range 0.0015-8 mM. 
The coefficient values were derived from diffusion times measured by FCS, (see 
Materials \& Methods).Both $D^{exp}_{146}(c)$ for DNA146 in very low salt 
$<0.05$ mM (sample set I) and for DNA146 in 10 mM TE buffer (sample set II) are 
shown. The concentration range of the FCS study encompasses the range studied by 
UV-spectroscopy. According to behavior of $D^{exp}_{146}(c)$ the studied 
concentration range has been divided into regions A,B,C,D with respective 
delimiting concentrations $c_{AB}=1$~mM, $c_{BC}=0.05$~mM and $c_{CD}=0.006$~mM.

In Fig.~\ref{diffcoef} we also show the diffusion coefficients $D^{ss}_{146}$ 
and $D^{ds}_{146}$, extrapolated for a DNA of a polymerization degree $N=146$. 
These are extrapolated (according to Eqs.\ref{DDCy5},\ref{extrapol}) from the 
diffusion times measured for 1.5 $\mu$M (basepair) of fluorescently labeled 
$N=110$ ssDNA and native dsDNA (DNA110*), respectively. First, to get 
$D^{ss}_{146}$ we used a sample of fluorescently labeled, single-stranded, 110 
nt long oligomer, in pure water solution. The diffusion time was extrapolated 
according to Eqs.\ref{DDCy5},\ref{extrapol} and we got $D^{ss}_{146}=3.95 \cdot 
10^{-11}$ m$^2$/s.  Then, to get $D^{ds}_{146}$ we used DNA110* added to 10 mM 
TE buffer, without any DNA146. Since in buffer, DNA110* is certain to keep the 
native, double helix conformation. Thus, $D^{ds}_{146}=3.2 \cdot 10^{-11}$ 
m$^2$/s is considered as the representative experimental value of the diffusion 
coefficient for native dsDNA, $N=146$.

First we note that, in region A ($c=1-8$ mM), the values $D^{exp}_{146}(c)$ for 
DNA146 both in 10 mM TE buffer and in very low salt $<0.05$~mM,  are equal 
within the experimental error. This result shows that DNA is of  native 
conformation in both environments. The variation of $D^{exp}_{146}$ is due to 
the  50 nm DNA146 fragments being in the semidilute regime above 1-2 mM \cite{TomicEPL08}.  
That is, the polyion diffusion slows down, due to an interchain repulsive 
interaction that increases with reduction of average distance between polyions, 
{\em i.e.} with the increase in concentration of polyions. Self-diffusion 
coefficient, $D_p(c)$ has been previously shown to decrease linearly with the volume fraction 
$\Phi$ of rod-like particles, for {\em e.g.} a mineral, boehmite, $L\approx 300$~nm, 
\cite{Dhont} and 20 bp dsDNA $L\approx 7$~nm, \cite{Wilk04}: 
\begin{equation} 
D_p(c)= D_p^0(1-\alpha \Phi) 
\label{Dhont} 
\end{equation} 
Here, $D_p^0$ is the self-diffusion coefficient measured in dilute solution. For 
DNA, molar concentration $c$ is related to volume fraction as $\Phi\approx0.001 
c$. Proportionality factor $\alpha$ depends on the aspect ratio $L/d$ of the 
particle \cite{Dhont}. For DNA146, $p\approx20-25$, leading to 
$\alpha\approx15$.  Fit of Eq.\ref{Dhont} to our DNA146 in TE buffer data gives 
$\alpha\approx30$ (full line in Fig.~\ref{diffcoef}), in a reasonable agreement 
with the theoretical value. Measurements at higher concentrations would be 
necessary to elucidate significance of this discrepancy, however this is out of 
scope here.

Below $c_{AB}=1$~mM (basepair), $D^{exp}_{146}(c)$, when measured in 10 mM TE is 
practically concentration independent across almost two decades in concentration 
(regions B,C). These values are very close both to just above defined 
experimental value for 146 bp dsDNA $D^{ds}_{146}=3.2 \cdot 10^{-11}$ m$^2$/s, 
and to the theoretical $D^{th}_{146}=3.27 \cdot 10^{-11}$ m$^2$/s value, 
calculated according to Eq.\ref{TiradoFormula} for a rodlike 146 bp dsDNA, 50nm 
long with $d=2.6$ nm diameter. That is, DNA146 in 10mM TE buffer keeps the 
native state. Also, since this is dilute regime, diffusion coefficient is 
constant. 

On the contrary,  $D^{exp}_{146}(c)$ in very low salt, deviates from the native 
DNA146 value, below $c_{AB}=1$~mM (basepair).  Specifically, in the intermediate 
concentration region B ($c=0.05-1$ mM) $D^{exp}_{146}=2.9 \cdot 10^{-11}$ 
m$^2$/s is rather constant, but below (beyond the experimental error) the native 
DNA146 value. This reduction in $D^{exp}_{146}(c)$ means that in region B, DNA 
polyions diffuse more slowly, than DNA polyions which keep the native 
conformation. 

Below $c_{BC}=0.05$~mM (basepair), in region C, $D^{exp}_{146}$ measured in very 
low salt starts to rise continuously and overpasses the native dsDNA146 value. 
Eventually, below $c_{CD}=0.006$~mM, in region D, $D^{exp}_{146}$ finally rises 
sharply to come very close to $D^{ss}_{146} =3.95 \cdot 10^{-11}$ m$^2$/s, 
the value that we ascribe to 146 nt ssDNA. Here we note that $D^{ss}_{146}$ 
value is higher than the theoretical value, $3 \cdot 10^{-11}$ m$^2$/s for a 
rodlike ssDNA that may be calculated according to Eq.\ref{TiradoFormula}, using 
$b=4.3$ nm and $d=1.2$ nm. Further, $D^{ss}_{146}$ value is lower than the 
experimental value $5 \cdot 10^{-11}$ m$^2$/s that may be extrapolated for 146 
nt ssDNA in the high salt (10 mM or higher) \cite{Tinland97}. In these high salt 
conditions ssDNA has a persistence length of the order of 1 nm and assumes a 
conformation of a compact, random walk coil \cite{OSF,Tinland97}. {\em E.g.}, 
146 nt ssDNA,  which has a contour length of about 60 nm, would form a coil with 
a hydrodynamic radius of 3 nm. Considering the hereabove presented range of 
diffusion coefficients for ssDNA, we judge that ssDNA in our, very low salt 
conditions neither assumes the  form of a compact coil, nor a rigid rod, but a  
rather extended coil.

\section{Discussion}
\label{sec4}
\subsection{Fraction of open basepairs}

As we have noted in previous Section, FCS measurements show that DNA146 in very 
low salt $<0.05$~mM (sample set I) assumes native, double-helix (and rodlike) 
conformation, but only in region A, above DNA concentration $c_{AB}=1$~mM. Also, 
we found that in region D, below  $c_{CD}=0.006$~mM  DNA appears to be 
single-stranded, forming a rather extended coil. In this manner, we recognize 
the concentration of DNA itself as a parameter to induce DNA melting in very low 
salt environment. 

Therefore we propose that our UV-absorbance data, that is, $c_{UV}/c$ ratio 
presented in Fig.\ref{UVgrav}, may be regarded as a DNA melting curve, recorded, 
however {\em vs.} DNA concentration as the parameter which induces melting. We 
remind that UV-absorbance only provides the amount of basepairs that have opened 
in a given DNA sample. Therefore, analysis of $c_{UV}/c$ data should provide the 
fraction $f(c)$ of open basepairs in the ensemble of DNA molecules in samples 
from set I. 

We also use the  Manning free  counterion fraction $\theta(c)$ data 
that we presented in Ref.~\cite{previous}, in support of $c_{UV}/c$ data. Both 
data sets have been obtained for sample set I. The reported $\theta(c)$ 
variation occurs within the range defined by the theoretical $\theta$ values for 
dsDNA and ssDNA. The relative variation that we obtain is also similar to the 
relative difference in $\theta$ values for dsDNA and ssDNA observed 
experimentally by other authors \cite{Auer69}. Thus, we consider that 
$\theta(c)$ curve represents DNA melting in low salt, and measures the fraction 
of open basepairs, as well as $c_{UV}/c$ data.

The open basepair fraction is derived in a following manner
\begin{equation}
f(c)= \frac{(A(c)-A_{min})}{(A_{max}-A_{min})}
\label{fc}
\end{equation}
Here $A_{min}$ and $A_{max}$ are the minimum and maximum value for the parameter
being analyzed, either $A(c)=c_{UV}/c$ or $A(c)=\theta (c)$.

\begin{figure}
\resizebox{0.44\textwidth}{!}{\includegraphics*{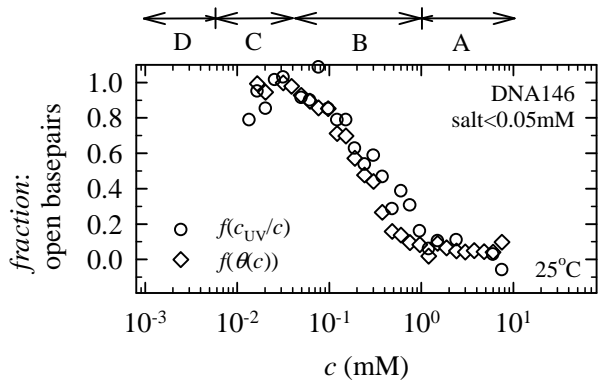}}
\caption{
The fraction $f$ of open basepairs {\em  vs.}~DNA146 basepair concentration. The 
circles are derived from $c_{UV}/c$ ratio (see also Fig.\ref{UVgrav}). For 
comparison we show data (diamonds) derived from free counterion parameter 
$\theta (c)$ taken from Ref.~\cite{previous}. Fraction $f\approx 0$ above 1 mM 
(region A) relates to dsDNA, while $f$ approaching the value of 1 below 
0.05 mM (region C) means that only ssDNA remains in the solution.
}
\label{openbp}
\end{figure}

Fig.~\ref{openbp} shows a very good correlation between the open basepair 
fraction  calculated from UV-absorbance $f(c_{UV}/c)$ and from free counterion 
fraction $f(\theta(c))$, across the studied DNA concentration range 
$c=0.015-8$~mM. We judge that the f(c) that we obtained is, indeed, a very god 
measure of fraction of open basepairs present in very low salt solutions of 
varying DNA146 concentrations. Both $c_{UV}/c$ and $\theta (c)$ data appear to 
be valid for characterization of the melting transition.

\subsection{Intermediate DNA}

The most important result of this study is that in region B the diffusion coefficient $D^{exp}_{146}$ measured for DNA146 in 
very low salt is diminished compared to values both for ssDNA and native dsDNA (DNA 
in 10 mM TE buffer). In this same region a 
fraction between 10 and 90\% basepairs are open ({\em cf.} Fig.\ref{openbp} and Fig.\ref{diffcoef}).

Careful consideration of the above leads to a conclusion that the DNA population 
in region B is NOT a simple mix of native dsDNA helix and separated ssDNA 
strands.  If it were a mix then the measured diffusion time and the 
corresponding diffusion coefficient should be of some intermediate value 
\cite{SchwilleBPC97}, between the experimental values for native dsDNA  and 
ssDNA, $D^{ds}_{146}=3.2 \cdot 10^{-11}$ m$^2$/s and $D^{ss}_{146}=3.95 \cdot 
10^{-11}$ m$^2$/s, respectively. However, the actual value is only $2.9 \cdot 
10^{-11}$ m$^2$/s! The DNA molecules there diffuse more slowly than either a 
rigid-rod dsDNA or an ssDNA strand. In other words, the DNA146 conformation 
present in this region is distinguished by being hydrodynamically larger entity 
than native DNA146 helix. As this is a dilute solution regime, the interactions 
between the molecules are not a plausible explanation, as they are for the 
decrease of $D^{exp}_{146}(c)$ above 1-2 mM, in the semidilute regime (region A, 
Fig.\ref{diffcoef}). 

Wilk \al~have also found a diffusion 
coefficient to be reduced for DNA fragments (20 bp)
in low salt, 0.01 mM NaCl in comparison to high salt, 0.2 M NaCl solutions  and for a rather similar DNA concentration range \cite{Wilk04}. However,  they have 
neglected in their analysis the fact that DNA denatures in the conditions of 
very low salt, which we have carefully analyzed in the previous subsection. To 
account for the reduction in diffusion ceofficient, they have proposed that the 
hydrodynamic volume of the DNA molecule is enlarged, by the amount of the Debye 
screening length $\kappa^{-1}$, that is, the contour length $L_c$ becomes 
$L_c+\kappa^{-1}$, and the diameter becomes $d+\kappa^{-1}$. We consider this 
scenario to be inappropriate. Already for DNA146 in 10 mM TE buffer (and a 
rather small $\kappa^{-1}=1$~nm), the value of the diffusion coefficient 
$D^{th}_{146}$ thus calculated (using Eq.\ref{TiradoFormula}) would deviate for 
10\% from the experimental result we got. We remind, we get the discrepancy of 
only 1-2\% between our experimental $D^{ds}_{146}$ result and theory, if we do 
not use a correction like Wilk {\em et al.}.  

Theoretically, the diffusion coefficient for rodlike molecules (see Eq.\ref{TiradoFormula}) is linearly dependent on the contour length of the molecule, 
and only logarithmically on its diameter. Thus, the most plausible mechanism for 
$D^{exp}_{146}(c)$ becoming reduced in region B would be if DNA contour length 
increased about 10\%, leading to a 10\% decrease in diffusion coefficient. Now, 
we note that dsDNA monomer size is 0.34 nm while for ssDNA it is 0.43 nm, 20\% 
longer \cite{Tinland97}. When a sequence of several basepairs in a dsDNA 
molecule breaks hydrogen bonds and unstacks (DNA bubble), then each strand 
locally assumes an ssDNA coil conformation.  Consequently the molecule would 
elongate in the region of the bubble. The molecules being elongated by 10\% - 
half the maximum 20\% - should then have half of the basepairs open. Simply, 
half of the sequence would be DNA bubbles and half native helix - an 
intermediate DNA would form. We assume that the fraying, separating ends of a 
dsDNA molecule would have a similar effect  as DNA bubbles, on slowing the 
intermediate DNA diffusion. 

Remaining issue would be whether these elongated, intermediate DNA  molecules  
really maintain rodlike shape, otherwise application of  Eq.\ref{TiradoFormula} 
would not be justified. We remind that, in these very low salt conditions, the 
charges along the DNA backbones repel strongly due to the lack of Debye 
screening and it is very plausible that this allows the intermediate DNA 
molecule to remain extended, rodlike in shape, despite having defects of lower intrinsic 
rigidity \cite{Yuan08,Lee10,TomicEPL08,Tinland97,Baumann97,TomicPRE07}.  We provide an 
illustration above the data panel of Fig.\ref{DNAconfo}.

\subsection{Fractions of DNA conformations}

\begin{figure}
\resizebox{0.44\textwidth}{!}{\includegraphics*{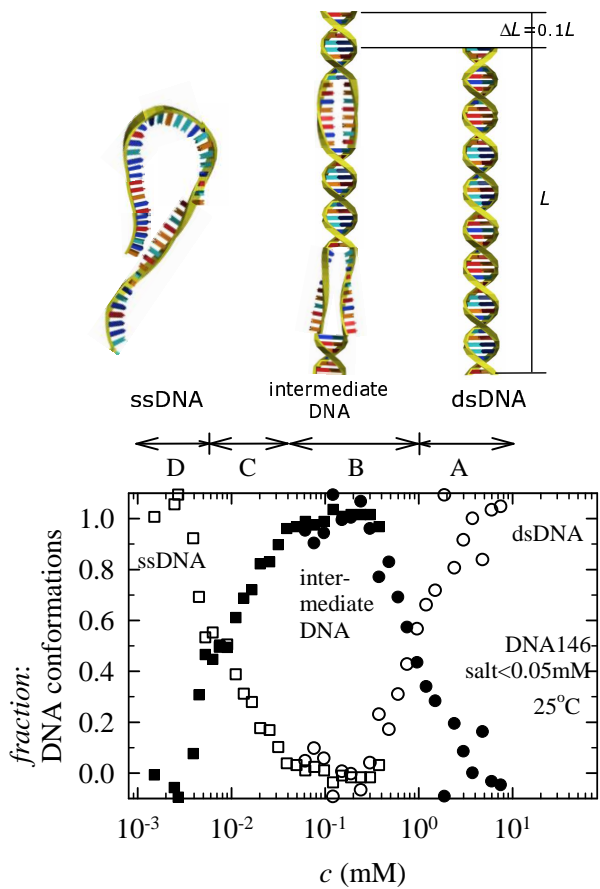}}
\caption{
Fraction of different conformations in the population of DNA146 molecules varies 
with DNA concentration (in basepair) in the very low salt ($<0.05$mM) solutions. 
Closed symbols  denote the intermediate DNA fraction $im(c)$. Open symbols 
denote the other conformation present in the population in the respective region 
($1-im(c)$). Circles and squares (both open and closed) denote fractions 
obtained according to Eq.\ref{Oc} \& \ref{ssc}, respectively. In region A, dsDNA 
dominates, starting to be replaced by intermediate DNA towards region B. Across 
most of the region B, intermediate DNA is dominant in the population. In region 
C  intermediate DNA  is replaced by ssDNA. Eventually, in region D, ssDNA 
dominates. Above the data-panel DNA conformations are sketched. The rodlike form 
and 10\% extension (compared to dsDNA) of intermediate DNA should be noted. 
}
\label{DNAconfo}
\end{figure}

For the analysis of fractions of conformations in the population of DNA146 molecules, 
we will postulate that intermediate DNA146 is of a rodlike form and is 
characterized with about 50\% basepairs open, and that the only other 
conformations that may appear in here analyzed solutions are fully separated 
ssDNA coils and rodlike native dsDNA helices. In other words, we do not expect 
DNA intermediate states with varying proportions of open base-pairs to coexist 
along the melting transition. Accordingly, we will analyze the diffusion 
coefficient data, and find how the fractions  of dsDNA helices, intermediate DNA 
and strands of ssDNA vary with the DNA concentration, in very low salt solution, 
Fig.\ref{DNAconfo}. Eventually, we will provide arguments to support this 
proposition.

First, we compare diffusion coefficient data obtained for DNA146 in very low 
salt  (set I) and in buffer (set II) in regions A and B. We denote these 
$D^{exp}_{146}(c)$ and $D^{ds}_{146}(c)$, respectively. Specifically, 
$D^{ds}_{146}(c)$, being obtained for DNA in buffer is thus the value for native 
helices, where the concentration dependence is due to samples above $c=1$~mM 
being in the semi-dilute regime. Variation in $D^{exp}_{146}(c)$ is also due to 
this influence, but, as DNA concentration is lowered below about 1-2 mM, 
$D^{exp}_{146}(c)$ deviates from $D^{ds}_{146}(c)$, showing that intermediate 
DNA starts to appear in the molecular population. The difference between 
$D^{exp}_{146}(c)$ and $D^{ds}_{146}(c)$ stabilizes in the lower part of region 
B, $c=0.05-0.3$~mM, indicating that only intermediate DNA remained in the 
population. We remind that the diffusion coefficient value $D^{im}_{146}=2.9 
\cdot 10^{-11}$ m$^2$/s in this region is lower than values either for dsDNA 
helices ($D^{ds}_{146}=3.2 \cdot 10^{-11}$ m$^2$/s) or ssDNA coils, and 
therefore can be representative only of intermediate DNA. Thus, the fraction of 
intermediate DNA $im(c)$ as compared with fraction $1-im(c)$ of native dsDNA across A and B regions is calculated in the following 
manner

\begin{equation}
im(c)=
\frac{(D^{exp}_{146}(c)-D^{ds}_{146}(c))}{(D^{im}_{146}-D^{ds}_{146})}
\label{Oc}
\end{equation}
In Fig.~\ref{DNAconfo} it is apparent that as DNA concentration is lowered dsDNA helices (open circles)
give way to intermediate DNA (black circles).

On lowering DNA concentration dsDNA disappears from population already in region 
B and only intermediate DNA and separated strands of ssDNA could remain in 
solution. The variation of $D^{exp}_{146}(c)$ across B,C,D regions may only be 
due to varying proportion of intermediate DNA and ssDNA coils, as they have 
different diffusion coefficients: $D^{im}_{146}=2.9 \cdot 10^{-11}$ and 
$D^{ss}_{146}=3.95 \cdot 10^{-11}$, respectively. Here DNA is in dilute regime 
and the diffusion coefficient of either conformation is not expected to depend 
on concentration. Thus, the fraction of 
intermediate DNA $im'(c)$ as compared with fraction $1-im'(c)$ of separated ssDNA strands across B,C,D regions is calculated in the following 
manner

\begin{equation}
im'(c)= \frac{(D^{exp}_{146}(c)-D^{im}_{146})}{(D^{ss}_{146}-D^{im}_{146})}
\label{ssc}
\end{equation}
In Fig.~\ref{DNAconfo} it is apparent that as DNA concentration is lowered 
intermediate DNA (black squares) separates into ssDNA coils (open squares), and in region D only ssDNA remains 
in solution. 

Finally, we note that $im'(c)$ is complementary with $im(c)$ data, and both 
indicate the same region $c=0.05-0.3$~mM where intermediate DNA is prevalent in 
the population. Interestingly, the fraction of open basepairs in this region 
is about 50\%, see Fig.\ref{openbp}. This supports our initial proposition that 
we do not expect DNAs with varying proportions of open base-pairs. 

Half of the basepairs being open for intermediate DNA relates to the fact that 
DNA that we used has a quite random sequence and equal content of G-C and A-T 
pairs. Since A-T pairs have smaller binding energy \cite{Breslauer86,Frank06}, 
it could be that only these break in the reported conditions and thus contribute 
to formation of intermediate DNA  as a step in DNA melting process.

Here, we remind that the counterion concentration is the only parameter that 
changes for this DNA and the reduction of screening due to decreasing 
concentration leads to DNA melting. Apparently, this parameter variation is mild 
enough to allow for preferential breaking of A-T pairs, or, which is rather 
equivalent \cite{Peyrard09EPL,Peyrard09JPCM}, formation of bubbles in AT-rich 
portions of the sequence.  That is, across our range of counterion 
concentrations 1-0.01 mM the Debye screening length $\kappa^{-1}$ varies from 
about 10 up to about 100 nm. The corresponding exponential factor $\exp(-\kappa 
r)$ in the screened Debye-H\"uckel potential at $r=2$nm, characteristic distance 
of two repulsing phosphate charges would concomittantly change for about 20\%. 
In other words, if the Coulombic repulsion at 1-2 mM counterion concentration 
was barely enough to break any basepairs, the reduction in screening at lower 
concentrations would enhance the repulsion for 20\% and start to break A-T pairs 
but not yet  G-C pairs which are bound 50-100\% stronger.

Considering the above illustration we note that Chen and Prohofsky \cite{Chen93}
calculate that below 1 mM added salt phosphate-phosphate force becomes
insensitive to the salt concentration, and conclude that the effective potential
has become a simple Coulomb potential.  However, this is a long range potential
and contributes to the energy balance of the intermediate molecule specifically
in the open, ssDNA sections, where the interaction occurs not only between
complementary bases, but among at least several neighboring bases. That is,
disrupted bond force constant is not zero. We emphasize that this renders the
calculation of free energy of such an intermediate state nontrivial.
Phosphate-phosphate contribution is non-additive, depends on DNA concentration,
and on the size of the bubbles. As such it is not tractable and not predicted by
present models for thermally induced DNA melting, where Coulomb potential was
neglected, being screened (at most, Coulomb potential has only renormalized the
hydrogen bonds).

\subsection{Concluding remarks}
In our scenario the intermediate DNA with about half basepairs 
open coexists with dsDNA helices or ssDNA coils in very low salt environment. We 
suggest that the low DNA concentration and very low salt conditions of this 
study allowed for DNA to open preferably along AT-rich portions of the sequence, 
without the strands separating further. These local lesions reflect on 
hydrodynamic properties of the whole molecule, {\em i.e.} its diffusion 
coefficient has been reduced measurably below the value for the original dsDNA 
helix.  However, whether these DNA lesions remain stably open or it is a 
dynamical situation where the bubbles open and close often enough to change the 
diffusion properties  may not be judged from our result. Experiments capable of 
measuring the fluctuation rates at a given site within the sequence, as those 
mentioned in the Introduction (optical spectroscopy of G bases, FRET, FCS) 
should be performed on DNA fragments in the very low added salt conditions. Only 
in this manner the nature of intermediate DNA could be judged and then also the 
stability mechanism  would be more easily inferred. An interesting scenario 
would be the critical slowing of fluctuations, a concept known from the studies 
of glass forming systems \cite{Lunken00}. Finally, our finding of intermediate 
DNA  provides another novel twist.  That is, to explain the diffusion properties 
of intermediate DNA in very low salt, the ssDNA loops within such a molecule 
have to be extended, and the molecule has to keep the rodlike form. This may be perceived contrary to the ususal notion that the
fluctuational openings, for DNA in an added salt environment, 
remove the molecule away from the rigid rod form \cite{Kafri00,Carlon02,Yuan08}.

\section{Summary} \label{sec5}

In this work we have revealed the presence of novel intermediate DNA 
conformation at low DNA concentrations ($c<1$~mM basepair) in very low salt 
($<0.05$mM) solutions at ambient temperature. The diffusion coeficient for 
DNA146 below 1 mM and above 0.05 mM (in DNA basepair) occurs to be below the 
values for both native dsDNA helices and ssDNA strands of equal polymerization 
degree. Thus the population in the intermediate regime is not a mixture of these 
two conformations. Intermediate DNA molecules, consisting of dsDNA segments and 
ssDNA loops would have appropriate hydrodynamical properties to explain for the 
diffusion data. The ssDNA loops are possibly due to DNA fluctuational openings 
(bubbles) that open and close at such rates, that they may be considered to have  become stabilized since they influence the diffusion of the whole molecule.   We suggest that the 
especially long-range Coulombic repulsion in the very low salt conditions of our 
experiment might allow for this effective stability.

\section*{Acknowledgement}

We gratefully acknowledge A.S. Smith and A. \v{S}iber for
illuminating
discussions and A.R. Bishop for inspiring this work. We thank A. Omerzu for kindly providing access to the
UV-spectrophotometer. This work is based on the support from the Unity through
Knowledge Fund, Croatia under Grant 22/08. The work was in part funded by
IntElBioMat ESF activity. The group at the Institute of physics works within
Project No. 035-0000000-2836 of Croatian Ministry of Science, Education and
Sports.


\begin{thebibliography}{0}


\bibitem{Polymerbooks}
 M.~Rubinstein, \and R.~H.~Colby {\em Polymer Physics}
{\em Oxford University Press, USA }(2003); J.~R.~C.~van der Maarel
{\em Introduction to Biopolymer Physics} {\em World Scientific,
Singapore} (2007)

\bibitem{Bloomfield00}
V.~A.~Bloomfield, D.~M.~Crothers and I.~Tinocco, Jr.,
{\em Nucleic Acids} (University Science Books, Sausalito, 2000).

\bibitem{Dove62}
W.~F.~Dove and N.~Davidson, J.~Mol.~Biol.,{\bf 5}, 467-478 (1962).

\bibitem{Record75}
M.~T.~Record, Jr., Biopolymers {\bf 14}, 2137 (1975).

\bibitem{Chen93}
Y.~Z.~Chen, and E.~W.~Prohofsky, Phys.~Rev.~E {\bf  48}, 3099
(1993).

\bibitem{Bishop}
M.~Peyrard and A.~R.~Bishop, Phys.~Rev.~Lett.~{\bf  62}, 2755
(1989); T.~Dauxois, M.~Peyrard, and A.~R.~Bishop, Phys.~Rev.~E {\bf  47}, 684
(1993).

\bibitem{Auer69}
H.~E.~Auer and Z.~Alexandrowicz, Biopolymers {\bf 8}, 1 (1969).

\bibitem{Breslauer86}
K.~J.~Breslauer, R.~Franks, H.~Blockers, and L.~A.~Marky,
Proc.~Natl.~Acad.~Sci.~USA {\bf 83}, 3746-3750 (1986).

\bibitem{Yakusevich04}
L.~V.~Yakushevich, {\em Nonlinear Physics of DNA, 2$^{nd}$ edition}, Wiley-VCH,
Weinheim (2004).

\bibitem{Frank87}
M.~D.~Frank-Kamenetskii, Nature, {\bf 328}, 17-18 (1987).



\bibitem{Gueron87}
M.~Gu\'{e}ron, M.~Kochoyan, and J.~L.~Leroy, Nature, {\bf 328}, 89-92 (1987).

\bibitem{Frank06}
A.~Krueger, E.~Protozanova, and M.~D.~Frank-Kamenetskii, Biophys.~J.~{\bf 90},
3091-3099 (2006).

\bibitem{Rapti06}
C.~H.~Choi, G.~Kalosakas, K.~\O.~Rasmussen, M.~Hiromura,A.~R.~Bishop, and
A.~Usheva, Nucl.~Acids.~Res.~{\bf 32}, 1584-1590 (2005);
Z.~Rapti, A.~Smerzi, K.~\O.~Rasmussen, A.~R.~Bishop,
C.~H.~Choi, and A.~Usheva, Europhys.~Lett., {\bf 74}, 540–546 (2006).

\bibitem{Manghi07}
M.~Manghi, J.~Palmeri and N.~Destainville, Phys.~Rev.~Lett.~{\bf 99}, 088103
(2007).

\bibitem{virus}
G.~D.~Pearson, J.~Virol., {\bf 16}, 17-26 (1975).

\bibitem{Bishop05}
S.~Ares, N.~K.~Voulgarakis, K.~\O.~Rasmussen, and A.~R.~Bishop,
Phys.~Rev.~Lett., {\bf 94}, 035504 (2005).


\bibitem{Peyrard09JPCM}
M.~Peyrard, S.~Cuesta L\'{o}pez and D.~Angelov, J.~Phys.: Condens.~Matter {\bf 
21}, 034103
(2009).

\bibitem{Scheraga66}
D.~Poland and H.~A.~Scheraga, J.~Chem.~Phys., {\bf 45}, 1456; {\bf 45}, 1464
(1966).


\bibitem{Manghi09}
M.~Manghi, J.~Palmeri and N.~Destainville, J.~Phys.: Condens.~Matter {\bf  21},
034104 (2009).

\bibitem{Kafri00}
Y.~Kafri, D.~Mukamel and L.~Peliti, Phys.~Rev.~Lett., {\bf 85}, 4988 (2000);
A.~Bar, Y.~Kafri, and D.~Mukamel, Phys.~Rev.~Lett., {\bf 98}, 038103 (2007);
H.~C.~Fogedby, and R.~Metzler, Phys.~Rev.~Lett., {\bf 98}, 070601 (2007).

\bibitem{Carlon02}
E.~Carlon, E.~Orlandini, A.~L.~Stella, Phys.~Rev.~Lett., {\bf 88}, 198101
(2002).



\bibitem{Gueron88}
J.~L.~Leroy, M.~Kochoyan, T.~Huynh-Dinh, and M.~Gu\'{e}ron, J.~Mol.~Biol.~{\bf
200}, 223–8 (1988).

\bibitem{Libchaber03}
G.~Altan-Bonnet, A.~Libchaber, and O.~Krichevsky, Phys.~Rev.~Lett., {\bf 90},
138101 (2003).


\bibitem{Peyrard09EPL}
S.~Cuesta L\'{o}pez, D.~Angelov and M.~Peyrard, Europhys.~Lett.~{\bf 87}, 48009
(2009).

\bibitem{Montrichok03}
Y.~Zeng, A.~Montrichok, and G.~Zocchi, Phys.~Rev.~Lett.~{\bf 91}, 148101 (2003);
A.~Montrichok, G.~Gruner and G.~Zocchi, Europhys.~Lett.~{\bf 62}, 452–8 (2003);
Y.~Zeng, A.~Montrichok, and G.~Zocchi, J.~Mol.~Biol.~{\bf 339},
67 (2004).

\bibitem{Wilcoxon83}
J.~Wilcoxon and J.~Michael Schurr, Biopolymers {\bf 22}, 2273-2321 (1983).

\bibitem{Yuan08}
C.~Yuan, H.~Chen, X.~W.~Lou, and L.~A.~Archer, Phys.~Rev.~Lett., {\bf 100},
018102 (2008).

\bibitem{Lee10}
O-chul Lee, Jae-Hyung Jeon, and Wokyung Sung, Phys.~Rev.~E {\bf 81},
021906 (2010).

\bibitem{Baumann97}
C.~G.~Baumann, S.~B.~Smith, V.~A.~Bloomfield and C.~Bustamante,
Proc.~Natl.~Acad.~Sci.~USA {\bf  94}, 6185 (1997).

\bibitem{TomicEPL08}
S.~Tomi\'{c}, S.~Dolanski Babi\'{c}, T.~Ivek, T.~Vuleti\'{c}, S.~Kr\v{c}a,
F.~Livolant, and R.~Podgornik,  Europhys.~Lett.~{\bf 81}, 68003 (2008).

\bibitem{ESOPS}
S.~Tomi{\'{c}}, D.~Grgi\v{c}in, T.~Ivek, S.~Dolanski Babi{\'{c}}, T.~Vuleti{\'{c}}, G.~Pabst, and R.~Podgornik, accepted in Macromol.~Symp.

\bibitem{OSF}
T.~Odijk, J.~Polym.~Sci.: Polym.~Phys. {\bf 15}, 477 (1977);
J.~Skolnick and M.~Fixman, Macromolecules {\bf 10}, 944 (1977).


\bibitem{previous}
T.~Vuleti{\'{c}}, S.~Dolanski Babi{\'{c}}, D.~Grgi\v{c}in, D.~Aumiler, J.~R\"adler,  F. Livolant and S.~Tomi{\'{c}}, submitted to Phys.~Rev.~E.

\bibitem{Mangenot03}
S.~Mangenot, S.~Keller and J.~R{\"{a}}edler, Biophys.~J.  {\bf 85}, 1817-1825
(2003).

\bibitem{ZeissManual}
Carl Zeiss: Applications Manual LSM 510 - ConfoCor 2 Application Handbook

\bibitem{Tirado84}
M.~Mercedes Tirado, C.~Lopez Martinez, J.~Garcia de la Torre, J.~Chem.~Phys. 81, 
2047-2051 (1984).

\bibitem{Stellwagen03}
E.~Stellwagen, Y.~Lu, N.~C.~Stellwagen, Biochemistry {bf 42}, 11745-11750 
(2003).

\bibitem{Manning72_a}
G.~S.~Manning, Biopolymers {\bf 11}, 937-949 (1972)

\bibitem{Benight}
R.~M.~Wartell and A.~S.~Benight, Phys.~Rep.~{\bf 126}, 67-107 (1985).

\bibitem{Buyuk}
 I.~E.~Scheffler, E.~L.~Elson, and R.~L.~Baldwin, J.~Mol.~Biol.~{\bf 48}, 145
(1970); S.~Buyukdagli and M.~Joyeux, Phys.~Rev.~E {\bf 78},
021917 (2007).

\bibitem{TomicPRE07}
S.~Tomi\'{c}, S.~Dolanski Babi\'{c}, T.~Vuleti\'{c}, S.~Kr\v{c}a,
D.~Ivankovi\'{c}, L.~Gripari\'{c} and R.~Podgornik,  Phys.~Rev.~E {\bf 75},
021905 (2007).



\bibitem{Dhont}
M.~P.~B.~Van Bruggen, H.~N.~W.~Lekkerkerker, and J.~K.~G.~Dhont, Phys.~Rev.~E {\bf 56},
4394 (1997);
J.~K.~G.~Dhont, M.~P.~B.~Van Bruggen, and W.~J.~Briels, Macromolecules
{\bf 32}, 3809 (1999).

\bibitem{Wilk04}
A.~Wilk, J.~Gapinski, A.~Patkowski, R.~Pecora,  J.~Chem.~Phys. {\bf  121}, 10794 (2004).


\bibitem{Tinland97}
B.~Tinland, A.~Pluen, J.~Sturm, and G.~Weill, Macromolecules {\bf 30}, 5763-5765
(1997).

\bibitem{SchwilleBPC97}
P.~Schwille, J.~Bieschke, F.~Oehlenschlager, Biophys.~Chem. {\bf 66}, 211-228
(1997).


\bibitem{Lunken00}
P.~Lunkenheimer, U.~Schneider, R.~Brand and A.~Loidl, Contemporary Physics {\bf 
41}, 15-36 (2000).

\end{thebibliography}
\end{document}